# Reproduction Experiments of Radial Pyroxene Chondrules Using a Gas-jet Levitation System under Reduced Conditions

Kana Watanabe , Tomoki Nakamura , and Tomoyo Morita
Department of Earth Science, Tohoku University, 6-3, Aramaki-Aza-Aoba, Aoba-ku, Sendai 980-8578, Japan

## Abstract

Reproduction experiments of radial pyroxene (RP) chondrules were carried out using an Ar–H$_2$ or Ar gas-jet levitation system in a reducing atmosphere in order to simulate chondrule formation in the protoplanetary disk. The experiments reproduced RP-chondrule textures, consisting of sets of thin pyroxene crystals and mesostasis glass between crystals. However, iron partition coefficients between pyroxene and glassy mesostasis (D$_{Fe}$ = Fe mol%$_{pyroxene}$/Fe mol%$_{mesostasis}$) in natural RP chondrules were much higher than that in the experimentally reproduced RP chondrules. The high D$_{Fe}$ in natural RP chondrules suggests that iron was removed from the mesostasis melt at high temperatures after the growth of pyroxene crystals. We found that many small iron-metal inclusions had formed in the mesostasis glass, indicating that FeO in the high-temperature melt of mesostasis was reduced to metallic iron, and iron in the mesostasis diffused into the newly formed metal inclusions. The formation of the iron-metal inclusions in the mesostasis was reproduced by our experiments in a reducing atmosphere, confirming that D$_{Fe}$ in natural RP chondrules increased after the growth of RP crystals. Therefore, the D$_{Fe}$ of RP chondrules can be an indicator to constrain cooling rates and redox states during chondrule formation.

## 1. Introduction

Chondrules are silicate spherules with diameters varying from several tens of micrometers to several centimeters (Prinz et al. 1988) and are the main components of chondrites. They formed by the melting of precursor dust particles at high temperatures and subsequent crystallization during cooling in the protoplanetary disk between 0.7 and 5 Myr after the formation of calcium–aluminum-rich inclusions (e.g., Kita & Ushikubo 2012; Bollard et al. 2015; Pape et al. 2019; Piralla et al. 2023). It has been suggested that chondrules interacted with ambient gas during formation based on the oxygen isotopic compositions and textures of olivine and pyroxene in porphyritic chondrules (Libourel & Portail 2018; Marrocchi et al. 2018, 2019; Piralla et al. 2021). Various chondrule formation models have been proposed, such as lightning models (Whipple 1966; Desch & Cuzzi 2000; Johansen & Okuzumi 2018), shock wave models (Connolly & Love 1998b; Ciesla & Hood 2002; Morris & Desch 2010), and planetesimal collision models (Kieffer 1975; Johnson et al. 2014, 2015; Wakita et al. 2017), but no widely accepted, definitive model exists at present. Constraints on the chondrule formation process lead to improved understanding of the temperature profiles and gas compositions of the protoplanetary disk.

Chondrule reproduction experiments have been carried out to understand the chondrule formation process. They revealed that the observed morphological characteristics of crystals in chondrules resulted from peak temperatures and cooling rates during chondrule crystallization (Tsuchiyama et al. 1980; Tsuchiyama & Nagahara 1981; Hewins 1983; Lofgren & Russell 1986; Lofgren 1989; Hewins & Radomsky 1990; Lofgren & Lanier 1990; Connolly et al. 1998a). However, in most of these experiments, the starting materials are fixed to Pt or Pt–Rh wires. The wire method is unsuitable for accurately reproducing the crystallization process in the protoplanetary disk because the contact points between the starting materials and wires work as heterogeneous nucleation sites and facilitate crystallization. For this reason, levitation experiments have been conducted to simulate chondrule formation (Tsukamoto et al. 1999; Nagashima et al. 2006, 2008; Pack et al. 2010; Mishra et al. 2020; Shete et al. 2021).

Chondrule formation took place under reducing conditions (e.g., Ebel & Grossman 2000; Grossman et al. 2008; Tenner et al. 2015; Villeneuve et al. 2015), and some of the following levitation experiments simulated those conditions. Mishra et al. (2020) conducted levitation experiments in a chamber filled with Ar gas under subatmospheric pressure and revealed that the latent heat of crystallization was removed more rapidly than under atmospheric conditions, resulting in faster cooling of the experimental samples that simulated chondrules. Consequently, this leads to smaller forsterite crystal sizes than under atmospheric conditions. The levitation experiments within an Ar gas–filled chamber by Shete et al. (2021) explored the relationship between the degree of supercooling at the time of enstatite crystallization and enstatite crystal volume in experimental samples. The starting materials used in these studies were pure forsterite or enstatite, while natural chondrules contain elements in addition to Mg, Si, and O. In particular, iron is a crucial element responsible for chondrule compositional variations, such as type I (Mg# > 90) and II (Mg# < 90) chondrules (Mg# = Mg/(Mg+Fe) × 100; Mcsween 1977; Scott & Taylor 1983). Pack et al. (2010)

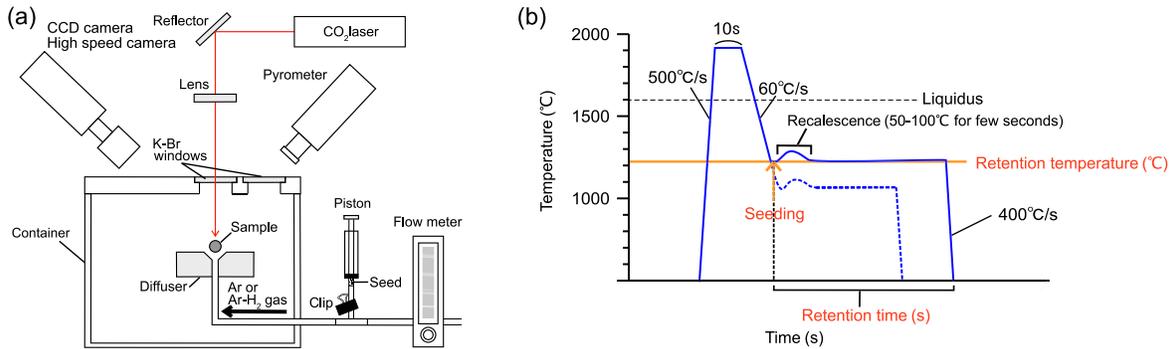

**Figure 1.** Overview of the experimental instrument and temperature profile. (a) Gas-jet levitator under the reducing conditions. A $CO_2$ laser (LC-100NV from DEOS) irradiated a levitated experimental sample, and the laser power was adjusted for heating and cooling. A reflector set the path of the laser beam, and a lens adjusted the beam to cover the entire sample. The sample was placed on a diffuser and floated by the gas flow ejected through a tube connected to the diffuser. A flow meter controlled the gas flow rate. In situ observations were conducted by a CCD camera (Elmo) and a high-speed monochrome camera (Photron FASTCAM-Net Max). The temperature was measured with a pyrometer (Lumasense Technologies Impac IN 140/5-H). The pyrometer measured the 5.14 $\mu$m wavelength in 500 °C–2500 °C, and the minimum measurement spot was 1.3 mm in diameter. The experiments were conducted in a container to control the redox state. A lid of the container had a valve opened during the experiments to allow gas to displace the inside air. K–Br windows on the lid transmitted the laser beam and emission from the sample for temperature measurements by the pyrometer. (b)Temperature profile during experiments. The solid blue line is an example, and the crystalized temperature, retention temperature, and retention time were changed as parameters along profiles such as the dashed line. The operation details are shown in Section 2.1.1

conducted levitation experiments and monitored the elemental evaporation behavior using silicates containing multiple elements under an Ar gas atmosphere. Their experiments also confirmed metal formation in experimental samples of ordinary chondrite composition under an Ar–$H_2$ atmosphere. However, they did not discuss the crystal morphologies and the elemental behavior between crystal and mesostasis glass during crystallization. Important conditions including cooling rate, redox state for establishing crystal morphologies, and chemical compositions of natural chondrules in experiments conducted under reducing conditions using natural chondrule compositions remain to be understood.

In this study, we conducted experiments to reproduce radial pyroxene (RP) chondrules using a gas-levitation system in a closed container under reducing conditions ($\Delta$IW $= -1.8$ to approximately WM (wustite-magnetite), $\Delta$IW (iron $-$ wustite) $\equiv \log(fO_2) - \log(fO_2)_{IW}$), which were close to the oxygen fugacities in the chondrule formation regions (Ebel & Grossman 2000; Tenner et al. 2015; Villeneuve et al. 2015). The starting materials had natural RP compositions, including iron. It is known that RP chondrules crystallized from totally melted precursor materials (Connolly & Hewins 1995) through heterogeneous nucleation by dust collisions (Nagashima et al. 2006). Levitation experiments are suitable for the reproduction of RP chondrules because our experimental procedure can simulate dust collisions with a floating melt sphere. In order to reproduce the heterogeneous nucleation of RP chondrules in this study, a small amount of powdered starting materials collided with a supercooled chondrule melt sphere as seeds to promote heterogeneous nucleation. This simulated the collisions with crystal nuclei such as interplanetary dust particles or small chondrules during chondrule formation in the protoplanetary disk. In the reducing conditions and with this particle collision procedure, we performed experiments with melting starting materials containing iron. Finally, we compared the crystal morphologies and chemical compositions of natural RP chondrules with those of our experimentally reproduced chondrules.

## 2. Methods

### 2.1. RP Chondrule Reproduction Experiments

#### 2.1.1. Experimental Procedure

The experimental chondrules were levitated on a gas diffuser at the Department of Earth Science at Tohoku University and heated or cooled by adjusting the power of a $CO_2$ laser (Figure 1(a)). The experiments were conducted in a container filled with Ar gas or Ar(97%)–$H_2$(3%) mixture gas. Before the experiments, the air in the container was replaced for 15–30 minutes by Ar gas or Ar(97%)–$H_2$(3%) gas, and the same gas was used for the levitation. Iron metal and magnetite appeared in the experimental chondrules, and iron metal indicates that reducing conditions were achieved in the container. The maximum oxygen fugacity in the container was approximately at the WM buffer, and the minimum one was calculated to be $\Delta$IW $= -1.8$, based on the iron mole fraction in the silicate (pyroxene $+$ mesostasis glass) and that in the metal in the experimental samples using the formula given in Corgne et al. (2008). The calculated oxygen fugacity indicates the condition at the retention temperature (Figure 1(b)), the temperature just before the experimental chondrules were quenched. The oxygen fugacity varied according to the time duration to replace air in the container by Ar–$H_2$ or Ar gas before the experiments (15–30 minutes) and the gas composition (Ar–$H_2$ or Ar) in the container. The magnitude of deviation from the IW buffer depended on the temperature at which a melt crystallized. The redox conditions in our experiments matched those in the chondrule formation regions where the oxygen fugacity ranged from approximately $\Delta$IW $= -4$ to $+1$ (Villeneuve et al. 2015).

The approximate temperature curve during the experiments is shown in Figure 1(b). First, a starting spherule was heated by a $CO_2$ laser at 1800 °C–2000 °C for 10 s, and we confirmed from the camera footage that the sample had melted completely. Subsequently, the melted sample was cooled at about 60°C s$^{-1}$ by adjusting the laser power and retained at a specific temperature below the liquidus (Table 1) for a few seconds to attain supercooling. At this temperature, we seeded the powder into the melted sample sphere. The seeds were powdered starting materials without sintering (mixtures of oxides of each element) or crushed crystals of experimental chondrules synthesized by the levitation system at the starting compositions. The oxide seed powders of the starting materials were 1–4 μm in diameter, and the crystal seed powders were adjusted to 42–52 μm by sieving with mesh sheets. The seed powder was sealed in a piston connected to a tube (Figure 1(a)) and introduced into the gas flow by opening a clip. About 10–20 seeds collided with the melted sample surface in each experiment. The samples crystallized at the seeding temperature, and, after crystallization, we maintained the samples at a temperature close to the seeding temperature for a set period of time (6–7200 s). The retention temperature is the average value during retention time and a little different from the seeding temperature because the temperature was adjusted manually to keep it constant while the temperature fluctuated slightly due to sample vibration by the gas flow. The laser was then turned off, and the sample was quenched. In this experiment, we could choose the crystallization temperature (equal to the seeding temperature), the retention temperature (almost equal to the crystallization temperature), and the retention time as parameters (Table 2). The experimental chondrules produced by this procedure were polished to make thin sections or polished mounts and observed by electron microscopes. The analysis conditions are shown in Section 2.2.2.

**Table 1**
Starting Material Compositions in Weight%

|  | RP-Fe1 | RP-Fe2 | RP-lowFe | RP-noFe |
|---|---|---|---|---|
| $Na_2O$ | 1.0 | 0.8 | 1.2 | 1.1 |
| MgO | 26.5 | 26.4 | 26.0 | 30.2 |
| $Al_2O_3$ | 2.5 | 2.3 | 2.6 | 3.0 |
| $SiO_2$ | 56.1 | 55.0 | 58.9 | 64.0 |
| CaO | 1.6 | 2.0 | 1.6 | 1.9 |
| FeO | 12.2 | 13.6 | 9.8 | 0.0 |
| Total | 99.9 | 100.1 | 100.1 | 100.2 |
| Liquidus (°C) | 1574.0 | 1574.0 | 1569.0 | 1597.0 |
| Emissivity | 0.65 | 0.65 | 0.66 | 0.80 |

### 2.1.2. Starting Material

We prepared four starting materials (Table 1) for the mixtures contained in RP chondrules. For iron, we used only FeO, and no metallic iron was included in the starting material. RP-Fe1 and RP-Fe2, almost identical chemical compositions, were close to the average compositions of natural RP chondrules calculated from the data of Dodd (1978), Lux et al. (1981), and Nagahara (1981). RP-lowFe is the bulk major-element composition of an RP chondrule in the Y-790448 LL3.2 chondrite measured in this study, and this composition contained a smaller amount of iron than RP-Fe1 and RP-Fe2. RP-noFe is a normalized composition of average RP chondrules excluding iron and trace elements. The liquidus temperatures of these starting compositions were calculated using rhyolite-MELTS (v.1.0.x.; Gualda et al. 2012), and those of RP-Fe1 and RP-Fe2 were 1574°C. The liquidus temperatures of RP-lowFe and RP-noFe were 1569 and 1597°C, respectively.

These starting materials were prepared as follows. First, the powdered starting material was placed on a carbon plate set in the container filled with Ar–$H_2$ or Ar gas and sintered by the $CO_2$ laser. Then the sintered material was floated on the diffuser by the gas flow and irradiated with the laser for about 10 s. The gases used in this step for each sample were the same ones that filled the container (Table 2). The materials melted and became a spherule due to surface tension. We used this sphere as a starting material. Subsequently, a series of levitation experiments were conducted with the spherule diameter ranging from 1.3 to 1.6 mm and the gas flow at 250–300 ml minute$^{-1}$. The list of experimental conditions and samples is shown in Table 2.

### 2.2. Natural Samples and Observation Procedures

#### 2.2.1. Meteorite Samples

To compare with the experimentally produced chondrule samples, 14 natural RP chondrules were chosen for the observation from chondrite meteorites that experienced the least aqueous alteration and thermal metamorphism because these secondary processes change the composition and mineralogy of chondrules: 6 chondrules from the Y-82038 (H3.2) chondrite (thin section 61-3), 7 from the Y-790448 (LL3.2) chondrite (thin sections 57-1 and 57-2), and one from the Y-81020 (CO3.0) chondrite (thin section 41-4). These thin sections were provided by the National Institute of Polar Research, Japan.

#### 2.2.2. Observation and Analysis Procedure

The natural and experimental chondrules were observed and analyzed by field-emission scanning electron microscopy (FE-SEM/EDS: JEOL JSM-7001F) at the Department of Earth Science at Tohoku University and the field-emission electron probe microanalyzer (FE-EPMA/WDS: JEOL JXA 8530F) at the Institute of Material Research at Tohoku University. The electron-beam acceleration voltage was 15 kV and the current was 1.0–1.4 nA for FE-SEM/EDS analysis, and 15 kV and 10 nA were used for FE-EPMA/WDS analysis.

### 3. Results

#### 3.1. Crystallization Processes and Structures of Chondrules

The levitation experiments indicated that, when the seed particles collided with the melt sphere, pyroxene crystallization started from the contact points of the seeds (Figure 2(a)). However, not all seeds remained as nucleation points. During the crystallization process, some seed particles melted in the melt sphere upon impact. Alternatively, the following possibilities can be considered. When crystallization began due to seed impact, the temperature of the experimental sample rose due to recalescence caused by the latent heat release. As a result, in some points with attached seeds, crystal growth became slower than in the initial crystallization point. The initial point, being denser than the remaining uncrystallized melt, caused the sample to levitate with the crystallized point at the bottom. Consequently, in this case, the melt (where crystallization was delayed) resided at the upper part of the sample and was melted by the laser irradiation from above, resulting in the initial crystallization regions serving as the sites for crystal growth. Low-Ca pyroxene was the only phase that crystallized. The crystal growth rates of the samples of RP-noFe composition, which were determined from the high-speed camera images, were very fast, about 0.2–1.3 mm s$^{-1}$, and the whole sample crystallized within about 8 s. The sample temperature increased

Table 2
Samples and Experimental Conditions

| Run No. | Starting Material | Gas | Seed | Crystallized Temperature (°C) | Retention Temperature (°C) | Retention Time (s) |
|---|---|---|---|---|---|---|
| 0710-1 | RP-lowFe | Ar | Oxide powder | 1063 | 1063 | 15 |
| 0710-3 | RP-lowFe | Ar | Oxide powder | 1063 | 1063 | 57 |
| 0728-2 | RP-lowFe | Ar–H$_2$ | Crystal powder | 1043 | 1052 | 100 |
| 0728-3 | RP-lowFe | Ar–H$_2$ | Crystal powder | 1057 | 1050 | 100 |
| 0728-4 | RP-lowFe | Ar–H$_2$ | Crystal powder | 1059 | 1066 | 100 |
| 0728-10 | RP-lowFe | Ar–H$_2$ | Crystal powder | 1040 | 1037 | 100 |
| 0728-14 | RP-lowFe | Ar–H$_2$ | Crystal powder | 1146 | 1138 | 100 |
| 0902-1 | RP-lowFe | Ar–H$_2$ | Oxide powder | 1241 | 1252 | 148 |
| 0902-4 | RP-lowFe | Ar–H$_2$ | Oxide powder | 1469 | 1477 | 21 |
| 0902-8 | RP-lowFe | Ar–H$_2$ | Oxide powder | 1552 | 1537 | 39 |
| 0902-10 | RP-lowFe | Ar–H$_2$ | Oxide powder | 1270 | 1257 | 15 |
| 0919-1 | RP-Fe1 | Ar–H$_2$ | Crystal powder | 1480 | 1294 | 20 |
| 0919-2 | RP-Fe1 | Ar–H$_2$ | Crystal powder | 1532 | 1411 | 21 |
| 0919-7 | RP-Fe1 | Ar–H$_2$ | Crystal powder | 1479 | 1483 | 48 |
| 0919-9 | RP-Fe1 | Ar–H$_2$ | Crystal powder | 1498 | 1536 | 3600 |
| 0919-11 | RP-Fe1 | Ar–H$_2$ | Crystal powder | 1480 | 1494 | 2400 |
| 1111-1 | RP-Fe1 | Ar | Crystal powder | 1422 | 1422 | 6 |
| 1111-4 | RP-Fe1 | Ar | Crystal powder | 1313 | 1313 | 24 |
| 1111-6 | RP-Fe1 | Ar | Crystal powder | 1274 | 1274 | 10 |
| 1111-7 | RP-Fe1 | Ar | Crystal powder | 1267 | 1267 | 10 |
| 1128-1 | RP-Fe1 | Ar–H$_2$ | No seeding | Not determined | 1549 | 4680 |
| 1128-2 | RP-Fe1 | Ar–H$_2$ | No seeding | Not determined | 1545 | 6000 |
| 1128-3 | RP-Fe1 | Ar–H$_2$ | No seeding | Not determined | 1505 | 2074 |
| 1128-6 | RP-Fe1 | Ar–H$_2$ | No seeding | Not determined | 1533 | 960 |
| 1128-7 | RP-Fe1 | Ar–H$_2$ | No seeding | Not determined | 1536 | 7200 |
| 1208-2 | RP-Fe2 | Ar–H$_2$ | No seeding | Not determined | 1430 | 523 |
| 1208-4 | RP-Fe2 | Ar–H$_2$ | No seeding | Not determined | 1396 | 1800 |
| 1208-5 | RP-Fe2 | Ar–H$_2$ | No seeding | 1460 | 1420 | 7200 |
| 0204-4 | RP-noFe | Air | Crystal powder | 1404 | 1404 | 12 |
| 0403-7 | RP-noFe | Air | Crystal powder | 1395 | 1395 | 9 |
| G7 | RP-noFe | Air | Crystal powder | 1397 | 1397 | 100 |

**Note.** Oxide powder = the powdered starting materials; crystal powder = crushed crystals of experimental chondrules with the same compositions as the samples.

by about 50 °C–100 °C for several seconds due to the recalescence caused by the latent heat release.

The chondrules produced by the experiments (Figures 2(c), (f)–(h)) were similar to the natural RP chondrules (Figures 2(b), (d), and (e)) in respect to crystal morphologies. The pyroxenes were plates or fine dendritic shapes and grew radially from several points of seed contacts on the sample surface. The crystal textures were different between samples kept at a temperature of crystallization for shorter times (<160 s) and those retained there for longer times (>500 s). In the samples with short retention times, Mg-rich pyroxene (hereafter "Mg-px") crystallized (Figure 2(f)). In contrast, relatively iron-rich pyroxene rims (hereafter "Fe-bearing px") were observed on the surface of early-crystallized Mg-px in the samples with long retention times (Figures 2(g) and (h)). Mesostasis glass was present in between pyroxene crystals. Iron metal inclusions and magnetite inclusions appeared in the mesostasis glass in the long-retained samples.

On the other hand, the natural RP chondrules in meteorites also consisted of plate shape pyroxenes and mesostasis glass, and the mesostasis contained small inclusions of iron–nickel metal and iron sulfides (Figures 2(d) and (e)). In some natural chondrules, the elemental distribution in the pyroxene crystals was almost homogeneous (Figure 2(d)), while in other samples, the edges of the pyroxene crystals (the brighter area in Figure 2(e)) were enriched in iron compared with the cores (the darker area in the same image), as was observed in pyroxene in the experimental RP chondrules (Figures 2(g) and (h)) with long retention times. The dendritic pyroxene, as observed in the experimental samples, was found in cryptocrystalline chondrules rather than typical RP chondrules.

### 3.2. Iron Partition Coefficient and Pyroxene Mg/(Mg+Fe) Ratios

Typical compositions of the natural and experimental RP chondrules measured by the electron microprobe are shown in Table 3. The listed analysis points were of the center of the mesostasis glass, the center of the pyroxene, and the zoned pyroxene rims. The pyroxene analyses (the cores and rims) are shown in the Supporting Data.

#### 3.2.1. Natural Chondrules

The iron partition coefficient $D_{Fe}$ is expressed as the ratio of iron mol% in pyroxene to that in mesostasis glass ($D_{Fe}$ = Fe mol%$_{pyroxene}$/Fe mol%$_{mesostasis}$). The $D_{Fe}$ of natural RP chondrules are plotted in Figure 3(a), which shows variable $D_{Fe}$ with an average of $D_{Fe}$ = 2.7, higher than 1. This indicates that pyroxene is more enriched in iron than mesostasis glass in natural RP chondrules.

Mg# of pyroxene in natural RP chondrules are also plotted in Figure 3(c), and the average Mg# is 77.0. Type II RP chondrules are common in the ordinary chondrites that we studied.

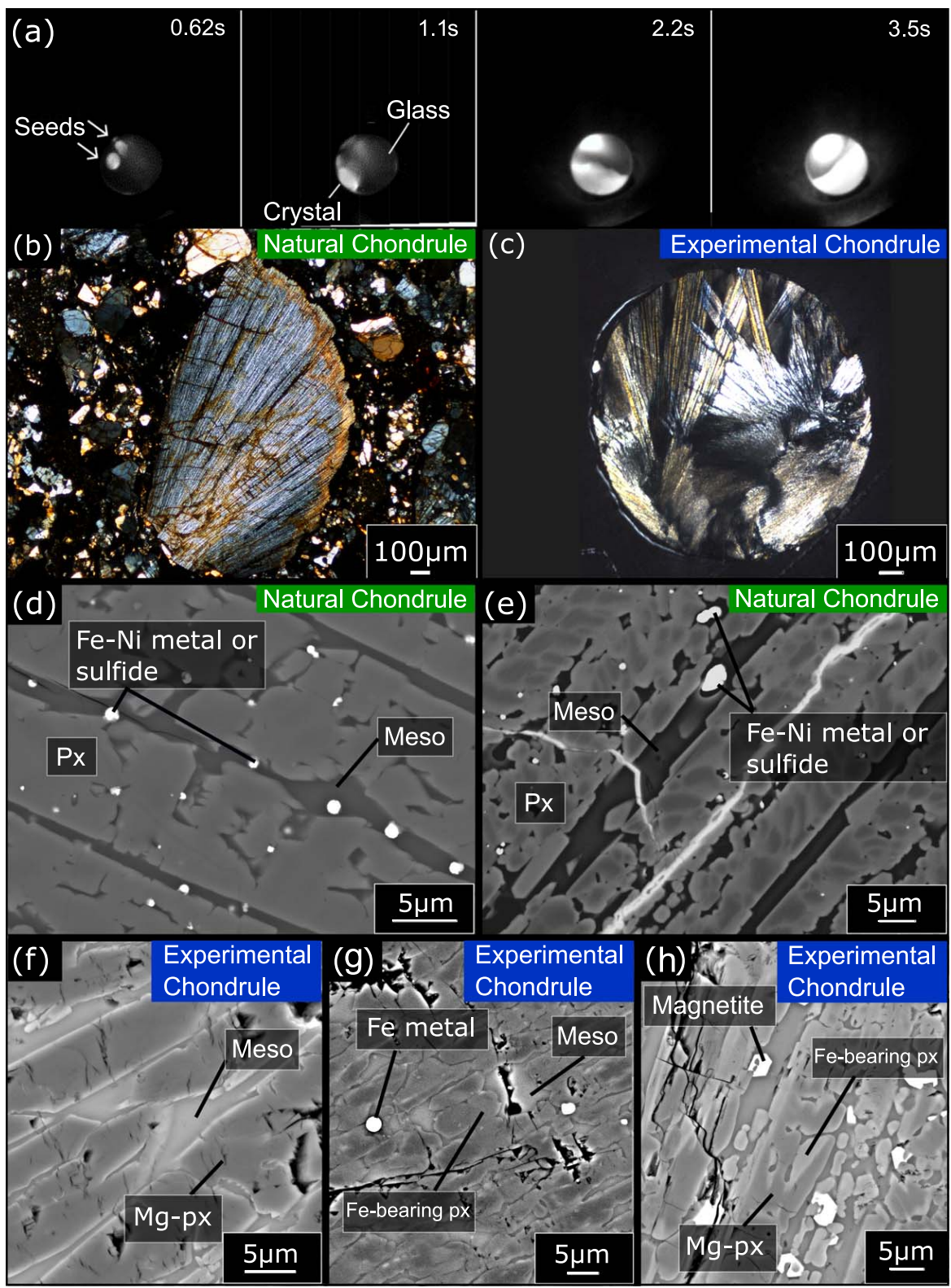

**Figure 2.** In situ observation and the textures of the natural and experimental RP chondrules. (a) Images from the high-speed camera during in situ observation. Brighter areas are crystals. The upper right values represent the time from the moment that the seeds were attached to the sample surface. The laser irradiated from above the sample, and the sample rotated during levitation. (b)–(c) Thin section images between crossed polarizers of (b) the RP chondrule in Y-790448(LL3.2), 57-2, and (c) the run G7 experimental chondrule. (d)–(h) Backscattered electron images of the natural and experimental chondrules: (d) RP chondrule in Y-790448(LL3.2), 57-1, thin section; (e) RP chondrule in Y-790448(LL3.2), 57-2, thin section; (f) run 0919-1 experimental chondrule; (g) run 0919-9 experimental chondrule; and (h) run 0919-11 experimental chondrule. Meso = mesostasis glass.

Table 3
Typical Analysis Compositions (Weight%) of Chondrules

|  | Natural Chondrules | | Experimental Chondrules | | | | |
|---|---|---|---|---|---|---|---|
|  | Y-790448, 57-1, No. 1 | | 0728-10 | | 0919-9 | | |
|  | Pyroxene | Meso | Pyroxene (Mg-px) | Meso | Pyroxene (Mg-px) | Pyroxene (Fe-bearing px) | Meso |
| $Na_2O$ | 0.19 | 2.09 | b.d. | 0.24 | b.d. | b.d. | 0.48 |
| MgO | 28.26 | 5.77 | 35.09 | 5.37 | 31.48 | 28.83 | 10.63 |
| $Al_2O_3$ | 1.4 | 10.03 | 0.37 | 16.01 | 1.14 | 2.13 | 12.23 |
| $SiO_2$ | 57.74 | 65.35 | 57.75 | 53.57 | 56.69 | 56.54 | 58.91 |
| $K_2O$ | b.d. | 0.72 | b.d. | b.d. | b.d. | b.d. | b.d. |
| CaO | 1.08 | 3.97 | 0.10 | 4.73 | 0.24 | 0.73 | 4.27 |
| $TiO_2$ | 0.12 | 0.25 | b.d. | b.d. | b.d. | b.d. | b.d. |
| MnO | 0.75 | 0.13 | b.d. | b.d. | b.d. | b.d. | b.d. |
| FeO | 10.71 | 3.05 | 6.49 | 17.53 | 9.45 | 12.97 | 12.19 |
| Total | 100.25 | 91.36 | 99.80 | 97.45 | 99.00 | 101.20 | 98.71 |

**Notes.** Meso = mesostasis glass; b.d. = below detection. The retention times of 0728-10 and 0919-9 are 100 and 3600 s, respectively.

### 3.2.2. Experimental Samples

The experimental chondrules showed lower $D_{Fe}$ than did natural RP chondrules (Figure 3(b)). When the retention times after crystallization were less than 150 s, the average $D_{Fe}$ was 0.4, indicating that the mesostasis was more iron-rich than pyroxene. The equilibrium $D_{Fe}$ between pyroxene and mesostasis glass was determined to be 0.37–0.65 in isothermal and slow cooling experiments of RP chondrules by Kennedy et al. (1993), whose starting compositions are relatively close to RP-Fe1 and RP-Fe2 in this study. $D_{Fe}$ of short (<150 s) retained samples in this study were included within the range of equilibrium values. Whereas, when the retention times were longer than 1000 s, the average $D_{Fe}$ was close to 1, and some exceeded 1. These high $D_{Fe}$ were observed between Fe-bearing px (the rim formed on Mg-px) and mesostasis glass. In other words, samples that retained longer times have $D_{Fe}$ closer to that of natural RP chondrules than those with short retention times.

The experimental samples showed higher Mg# than did the natural RP chondrules (Figure 3(d)). The Mg# of Mg-px with short retention times (<160 s) were approximately 90 regardless of starting compositions. Meanwhile, the long retention times (>500 s), Mg# of Fe-bearing px (the rim on Mg-px) were lower and close to that of natural RP chondrules (the average Mg# of natural RP was 77.0). Mg# of Mg-px, inside Fe-bearing px (the cores of pyroxene), were approximately 90, similar to Mg-px in the samples with short retention times. The rims of pyroxene in some of the natural RP chondrules were also more enriched in iron than the cores, like pyroxenes in the experimental samples (Figure 2(e)). The gas composition (Ar–$H_2$ or Ar) used in the experiments did not affect $D_{Fe}$ and Mg# (Figures 4(a) and (b)). The plotted values in Figure 4 are shown in Supporting Data.

## 4. Discussion

### 4.1. Constraints on RP Chondrule Formation from Iron Partition Coefficients and Elemental Distribution

#### 4.1.1. High Iron Partition Coefficients in Natural RP Chondrules

In the experimental samples of Kennedy et al. (1993) and experimental chondrules in this study with short retention times after crystallization (<150 s), $D_{Fe}$ were as low as 0.3–0.7 (Figure 3(b)). In contrast, natural RP chondrules show high $D_{Fe}$, well above 1 ("original" values in Figure 3(a)), and the experimental chondrules that were retained for a long time after crystallization (>1000 s) showed $D_{Fe}$ that were close to or exceeded 1 (Figure 3(b)). The $D_{Fe}$ difference is important because, in crystallization, whether $D_{Fe}$ are less than or larger than 1 reflects more iron being distributed in the crystal or the mesostasis melt. Note that in this study, $D_{Fe}$ of natural and experimental chondrules were defined as being established after chondrule solidification, not during crystallization.

The $D_{Fe}$ similarity of natural chondrules and the experimental samples with long retention times indicates that both experienced the same process during formation and thus constrains the formation process of natural RP chondrules.

First, we consider why $D_{Fe}$ approached 1 in the experimental chondrules with increasing retention times. The first possible cause is the growth of Fe-bearing px. As Fe-bearing px grew, which contained more iron than the underlying Mg-px, the iron content in the mesostasis glass decreased, because iron was supplied by mesostasis glass, not by Mg-px. As a result, $D_{Fe}$ increased and approached 1 ($D_{Fe}$ were defined between Fe-bearing px and mesostasis).

The second possible cause is the reduction of FeO to metallic iron or the oxidation of FeO to magnetite in the mesostasis melt due to the long retention times in $\Delta IW = -1.8$ to an approximately WM atmosphere (Figures 2(g) and (h)). In most cases, the samples with long retention times had metals or magnetites in mesostasis glass. In particular, the occurrences of spherical iron metal inclusions in the experimental chondrules are similar to iron–nickel metal and sulfides observed in natural RP chondrules (Figures 2(d) and (e)). Mesostasis glass near the metal or magnetite boundary is darker in the backscattered electron image (Figures 2(g) and (h)). This means that iron is more depleted than the average mesostasis glass, suggesting that the metals or magnetites nucleated in the mesostasis melt after pyroxene growth. Since the valence of iron in the starting material was only divalent (FeO), the divalent iron in the mesostasis was reduced or oxidized, and metals or magnetites were formed before the solidification of mesostasis melt into the glass. When the metals or magnetites crystallized in the mesostasis melt, the reduced or oxidized iron in the mesostasis was absorbed by (concentrated into) the metal or magnetite crystals, and thus the iron content in the mesostasis was

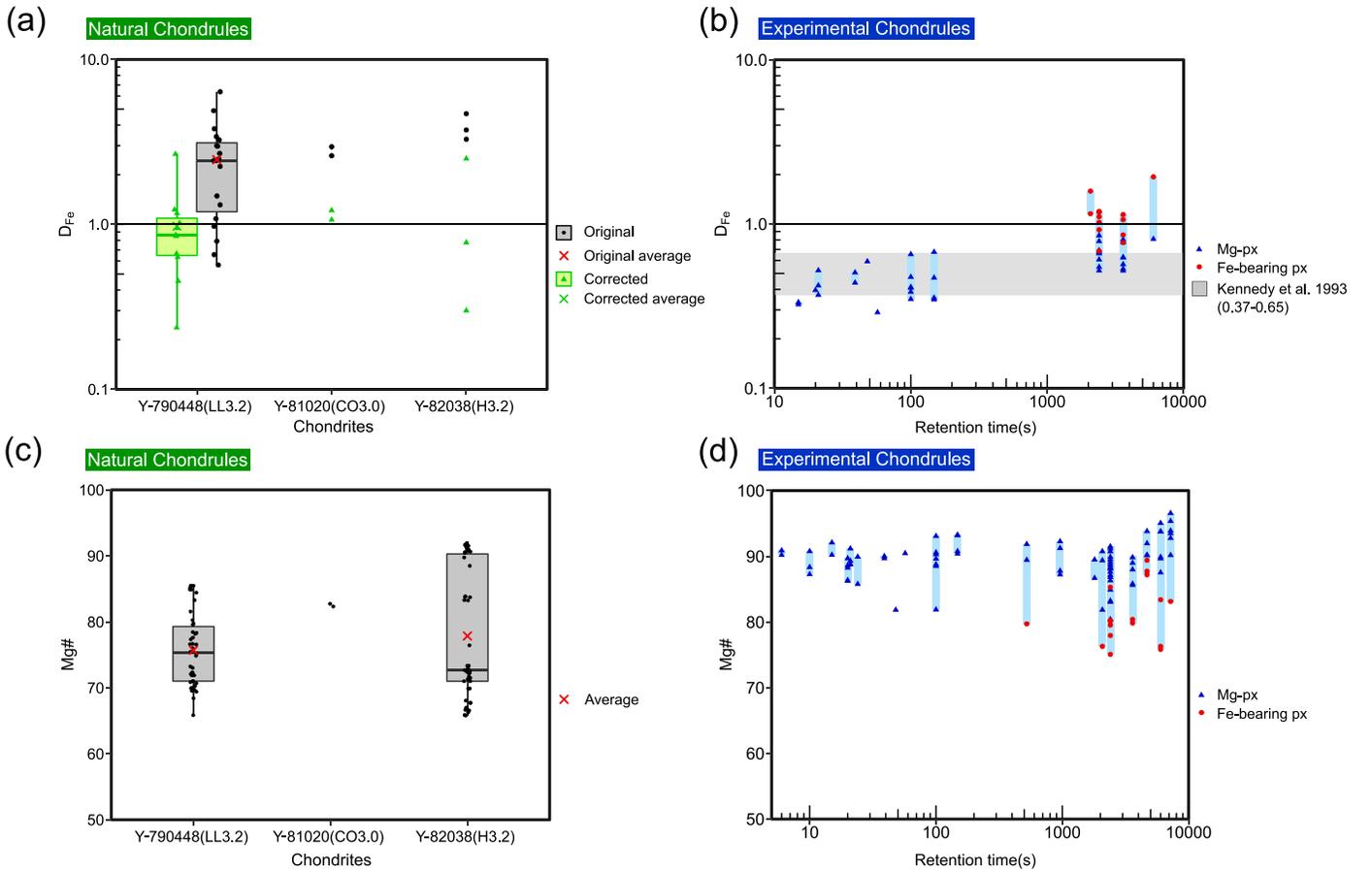

**Figure 3.** Comparison of the natural and experimental chondrules. (a) Box plot of $D_{Fe}$ of natural chondrules with interquartile range and whiskers from the minimum to the maximum value. "Original" is the analyzed value. "Corrected" is the $D_{Fe}$ derived from iron in pyroxene and that in whole mesostasis (mesostasis glass including metals and sulfides). (b) $D_{Fe}$ in experimental chondrules. Mg# in (c) natural chondrules (box plot) and (d) experimental. Blue lines in (b) and (d) represent the minimum to maximum value range. In (b) and (d), Mg-px and Fe-bearing px regions overlap in some places. Because we define them in a single crystal, the Mg-px in some crystals are iron-rich relative to Fe-bearing px in other crystals.

(The data used to create this figure are available.)

reduced. As a result, $D_{Fe}$ between pyroxene and mesostasis glass increased.

A similar process can be considered for natural RP chondrules. Iron–nickel metals and sulfides are commonly distributed in the mesostasis glass of natural RP chondrules (Figures 2(d) and (e)). After pyroxene growth, the FeO in the mesostasis melt might have been reduced or sulfurized and the iron metals or iron sulfides nucleated in the melt. As a result, FeO content in the mesostasis decreased, and the high $D_{Fe}$ were established in natural RP chondrules. "Corrected $D_{Fe}$," which is recalculated $D_{Fe}$ between pyroxene and whole mesostasis (total iron concentration in mesostasis glass including metal and sulfide inclusions), are plotted around 1 (Figure 3(a)). Therefore, $D_{Fe}$ in natural RP chondrules should be "apparent $D_{Fe}$" that were established after pyroxene growth and do not represent iron distribution during crystal growth. $D_{Fe}$ in the experimental chondrules containing iron metals are lower compared to natural RP chondrules (Figures 3(a) and (b)), possibly because longer retention times or more reducing gas in the protoplanetary disk facilitated the formation of iron metal inclusions, which resulted in the increase in $D_{Fe}$.

Libourel & Portail (2018), Marrocchi et al. (2018, 2019, 2022), and Piralla et al. (2021) suggested that gas–melt interaction occurred during chondrule formation. According to them, chondrules that experienced gas–melt interactions exhibit asymmetric crystal morphologies or elemental composition gradients like zoning or overgrowth layers in crystals. If natural RP chondrules underwent gas–melt interactions, they should also exhibit such characteristics in pyroxene crystals, and $D_{Fe}$ will be affected. However, the natural RP chondrules observed in this study (Figure 2(d)) did not clearly display the pyroxene morphologies and elemental composition gradients suggested for gas–melt interactions. Some natural RP chondrules, like Figure 2(e), had iron-rich rims in pyroxene; however, if these rims were the result of gas–melt interaction between iron-rich gas and chondrule melt, both the crystal and the mesostasis glass would become iron-rich as iron-rich gas diffused into the chondrule melt during crystallization. Therefore, the coexistence of iron-rich pyroxene and iron-poor mesostasis glass $D_{Fe}$ that exceed 1, observed in this study, cannot be explained by gas–melt interactions, and a more suitable explanation is iron removal from mesostasis melt by the formation of metals or sulfides.

*4.1.2. Pyroxene Composition in Natural and Experimental RP Chondrules*

Natural RP chondrules, at least in the ordinary chondrites we studied, are commonly classified as type II based on their pyroxene compositions. In our experiments with short retention times (<160 s), Mg-px (Figure 2(f)) crystallized and showed higher Mg# (the average was approximately 90) compared with that in natural chondrules. In contrast, Fe-bearing px (Figures 2(g) and (h)) that crystallized with long retention times (>500 s) had Mg# close to that in natural RP chondrules. The long retention times after crystallization simulate slow cooling after crystallization. Thus, natural RP chondrules might have cooled sufficiently slowly after the crystallization for Fe-bearing px to grow.

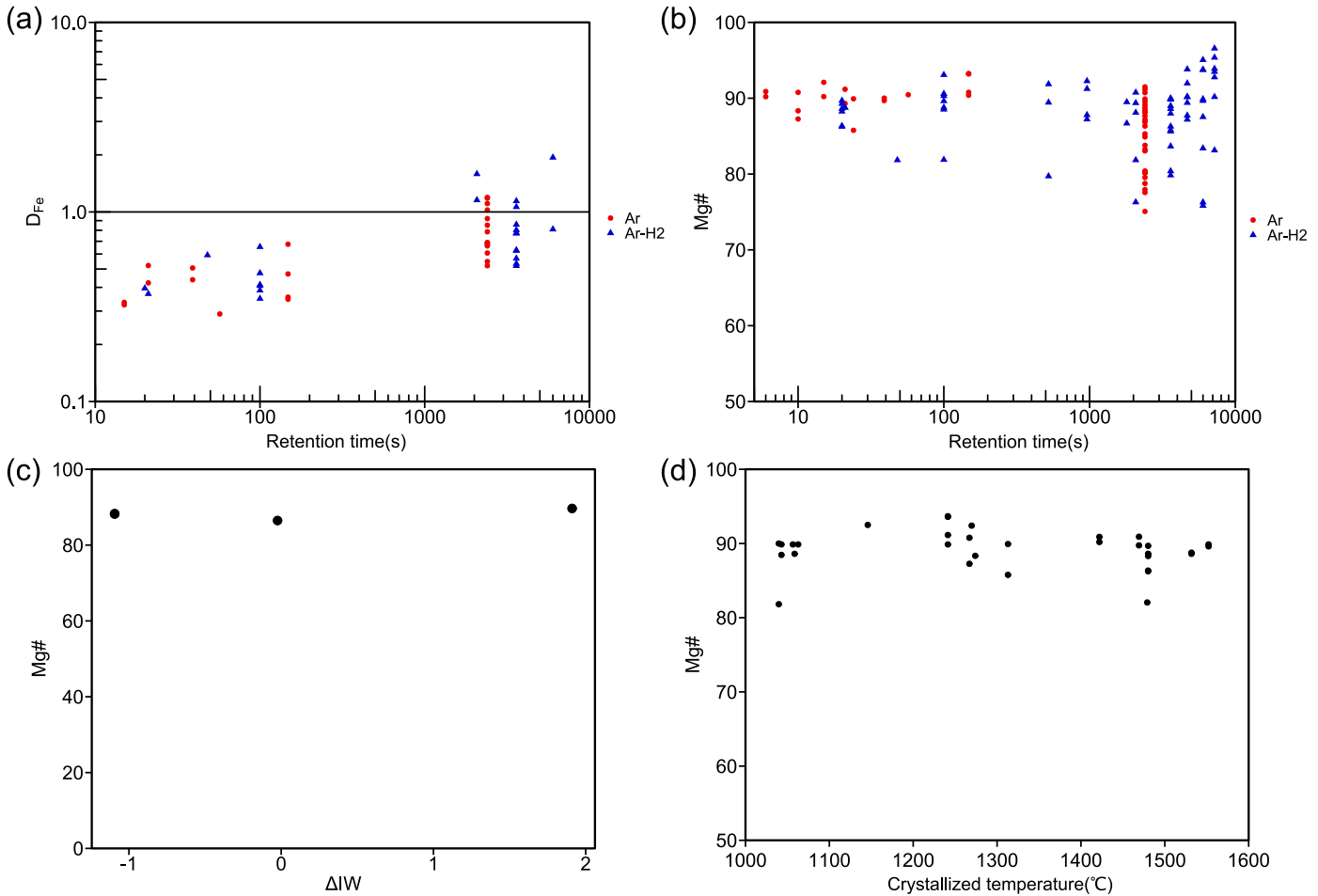

**Figure 4.** Relationships between gas used in experiments and (a) $D_{Fe}$ and (b) pyroxene Mg# and relationships between Mg# and (c) oxygen fugacity and (d) crystallization temperature.
(The data used to create this figure are available.)

In the experimental chondrules with long retention times, the occurrence of Fe-bearing px is considered to be the result of either overgrowth on earlier formed Mg-px or iron diffusion from mesostasis to the rim of Mg-px. In particular, the textures of the overgrowth are remarkable: the sharp boundaries between Mg-px and Fe-bearing px and structures where more than two pyroxene plates fused by Fe-bearing px (Figure 2(g) and (h)). The overgrowth and/or iron diffusion can explain the low pyroxene Mg# in natural RP chondrules.

We investigated the possibility that the oxygen fugacity during crystallization may change the ion valence and affect the pyroxene Mg#. Our experiments suggest that there were no relationships between pyroxene Mg# and the oxygen fugacity controlled by gas species composition (Ar–$H_2$ or Ar;

Figure 4(c)). In addition, the crystallization temperature also did not affect pyroxene Mg# (Figure 4(d)). The plotted values in Figure 4 are shown in Supporting Data.

Therefore, it is suggested that natural RP chondrules were produced by pyroxene crystallization followed by cooling slow enough to allow the crystallization of metal and sulfide inclusions in the mesostasis melt, diffusion of iron from mesostasis into pyroxene, and overgrowth of Fe-bearing px on earlier formed Mg-px. Additional experimental studies of the reduction and sulfidation of mesostasis, the effects of the cooling rates of chondrules, and the ambient gas conditions would further constrain chondrule formation events.

### 4.2. Crystallization of RP Chondrules by Seeding

The high-speed camera images (Figure 2(a)) showed that RP crystallized by heterogeneous nucleation from the seed contact points. In the protoplanetary disk, interplanetary dust particles could collide with chondrule melts as the seeds. The collisions are possible due to the high dust densities in the chondrule-forming regions (Ebel & Grossman 2000; Fedkin & Grossman 2013). If we suppose that RP chondrules crystallized by collisions with dust grains, the number density of dust grains that existed during chondrule formation events can be estimated. For instance, the mass density of the chondrule itself and the chondrule mass density in chondrule-forming regions are assumed to be $3.0\,\mathrm{g\,cm^{-3}}$ (Ciesla et al. 2004) and $6\times10^{-12}\,\mathrm{g\,cm^{-3}}$ (Arakawa & Nakamoto 2019), respectively, and the chondrule diameter can be set to 0.5 mm. The chondrule number density of $3.1\times10^{-8}\,\mathrm{cm^{-3}}$ is given from these values, and the number density of the dust grains is assumed to be higher than that of chondrules because not all dust collides with chondrules.

## 5. Conclusions

Heating experiments of RP chondrules in a reducing atmosphere with natural RP chondrule compositions reproduced the radial texture of low-Ca pyroxene crystals and interstitial mesostasis glass. Comparison between experimentally reproduced chondrules and natural chondrules indicates that the high iron partition coefficients between pyroxene and mesostasis glass in natural RP chondrules are explained by the process of iron removal from the mesostasis due to the formation of the metal inclusions in the mesostasis melt. The growth of iron-rich secondary pyroxene partly explains the variation of the Mg# of pyroxene in natural RP chondrules during slow cooling. Natural RP chondrules experienced slow cooling that allowed these processes to occur.

## Acknowledgments

We would like to thank Dr. K. Tsukamoto, Dr. T. Yoshizaki, Dr. A. Tsuchiyama, and Dr. Y. Kimura for their valuable comments and suggestions, which greatly improved our manuscript. We would also like to thank Dr. M. Zolensky for the English improvement of the manuscript. The National Institute of Polar Research, Japan, provided the thin sections of chondrites.

## ORCID iDs

Kana Watanabe https://orcid.org/0009-0008-8397-6119
Tomoki Nakamura https://orcid.org/0000-0001-9525-4086
Tomoyo Morita https://orcid.org/0009-0002-2769-3567